\documentclass{icrc2009}

\usepackage{graphicx}
\usepackage[caption=false]{caption}
\usepackage[font=footnotesize]{subfig}
\usepackage{fixltx2e}
\usepackage{url}
\usepackage{multirow}

\newcommand{\shorttitle}[1]%
{\markboth{Proceedings of the 31\MakeLowercase{$^{st}$} ICRC, {\L}\'{o}d\'{z} 2009}{#1} }
\newcommand{\etal}{\MakeLowercase{\textit{et al. }}}

\begin{document}
\title{Blazar Discoveries with VERITAS}

\author{\IEEEauthorblockN{J. S. Perkins\IEEEauthorrefmark{1}
    for the VERITAS Collaboration\IEEEauthorrefmark{2}}\\
  \IEEEauthorblockA{\IEEEauthorrefmark{1}Fred Lawrence Whipple
    Observatory, Harvard-Smithsonian Center for Astrophysics, Amado,
    AZ 85645, USA \\ (jperkins@cfa.harvard.edu)}
  \IEEEauthorblockA{\IEEEauthorrefmark{2}see R.~A.~Ong
    \etal\cite{ong.2009} or
    http://veritas.sao.arizona.edu/conferences/authors?icrc2009}}

\shorttitle{J.S.Perkins \etal, VERITAS Blazar Discoveries}
\maketitle

\begin{abstract}
  Blazars are among the most energetic and violent objects in the
  universe. By observing blazars at very high energies (VHE, E $>$ 100
  GeV) we can better understand blazar emission mechanisms, jet
  structure, black hole accretion and the extragalactic background
  light (EBL). VERITAS, an array of four 12-meter diameter imaging
  atmospheric-Cherenkov telescopes, performs VHE studies of blazars
  through intense monitoring and discovery campaigns. Most blazars
  known to emit VHE gamma rays are high-frequency-peaked BL Lacertae
  (HBL) objects, and VERITAS has discovered VHE emission from two of
  these: 1ES 0806+524 and RGB J0710+591. VERITAS has also discovered
  VHE gamma rays from two intermediate-frequency-peaked BL Lacertae
  (IBL) objects: W Com and 3C 66A. The expansion of the VHE catalog
  to include IBL objects enables a better understanding of the AGN
  population as a whole. This contribution presents recent results
  from the VERITAS blazar discovery program.
  \end{abstract}

 
\section{Introduction}

VERITAS is an array of four 12m diameter imaging atmospheric Cherenkov
telescopes (IACT) located at the Fred Lawrence Whipple Observatory in
southern Arizona at an elevation of 1268m \cite{weekes.2002}.  In less
than 50 hours of observations, VERITAS can detect a source at the 1\%
Crab flux level in the energy range from 100 GeV to greater than 30
TeV.  The array can reconstruct events with an energy resolution of
$\sim$15\% and an angular resolution (68\% containment) of
$\sim$0.1$^\circ$.  For more details on the VERITAS instrument and
techniques, see \cite{ong.2009}.  The VERITAS capabilities are well
suited to the study and search for blazars in the Northern Hemisphere
and will become further enhanced once a planned move of one of the
telescopes is completed later this year which will yield a $\sim$15\%
improvement in sensitivity.

\begin{figure}[!t]
  \centering
  \includegraphics[width=2.5in]{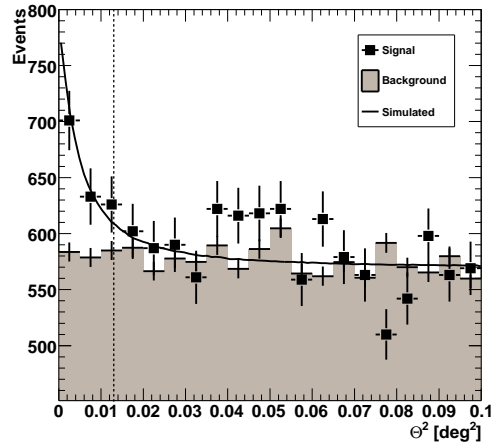}
  \caption{Distribution of the squared angular distance between the
    shower direction and the location of 1ES 0806+524.  The vertical
    dashed line indicates the size of the integration region.  The
    solid line indicates the expected shape of the distribution for a
    point source.}
  \label{fig:0806theta2}
\end{figure} 

A large fraction of the $\sim$750 hours of dark time (plus an
additional $\sim$250 hours of moon time) during the VERITAS observing
season is dedicated to observing active galactic nuclei (AGN).  The
focus of these observations is two fold: one, to study known VHE AGN
such as Mrk421 or 1ES 1218+304 to learn about blazar emission
mechanisms, black hole accretion and the extragalactic background
light (EBL) and, two, to discover new types of VHE blazars which will
expand the current VHE blazar catalog and perhaps reveal unknown VHE
phenomenon. For more details about the VERITAS blazar key science
program, see W.~Benbow's contribution in this
conference\cite{benbow.2009}.

\begin{figure}[!t]
  \centering
  \includegraphics[width=2.5in]{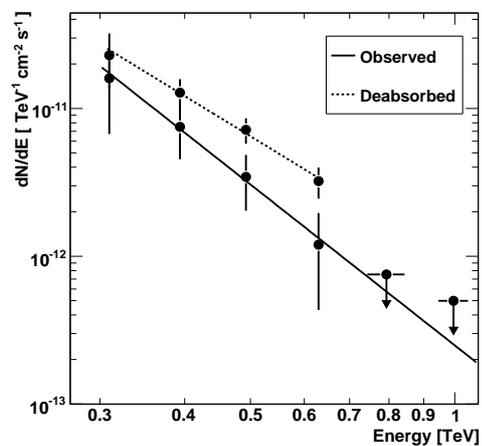}
  \caption{Differential photon spectrum of 1ES 0806+524.  The spectrum
    is well fit by a power law with index $3.6 \pm 1.0_{stat} \pm
    0.3_{sys}$.  The de-absorbed spectrum is calculated by applying
    the extragalactic absorption model according to Franceschini
    \etal, 2008\cite{franceschini.2008}.}
  \label{fig:0806spectrum}
\end{figure}

\begin{figure*}[!t]
  \centerline{\subfloat[W Com Skymap]{\includegraphics[width=2.8in]{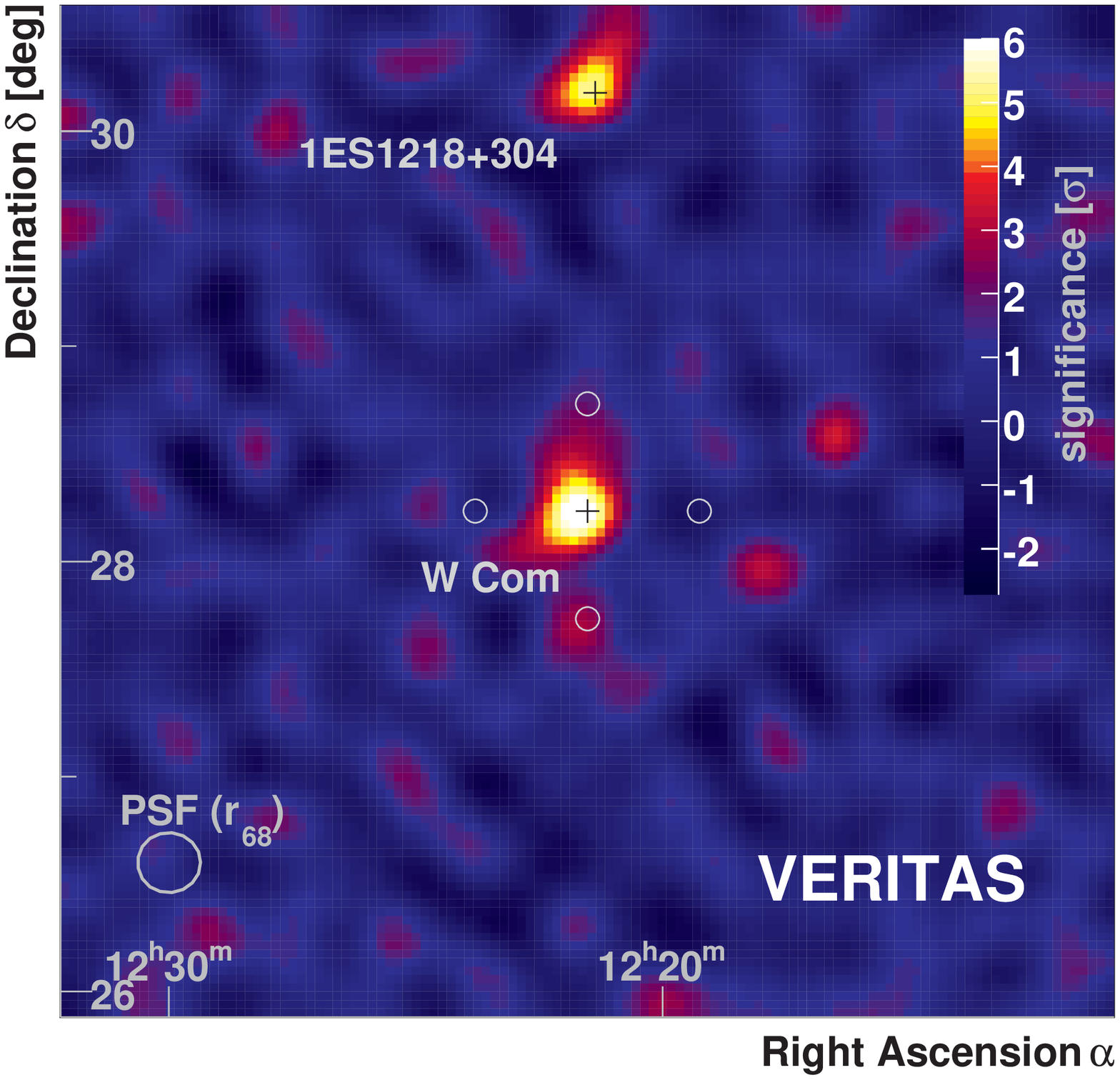} \label{fig:wcomskymap}}
    \hfil
    \subfloat[W Com Lightcurve]{\includegraphics[width=2.2in]{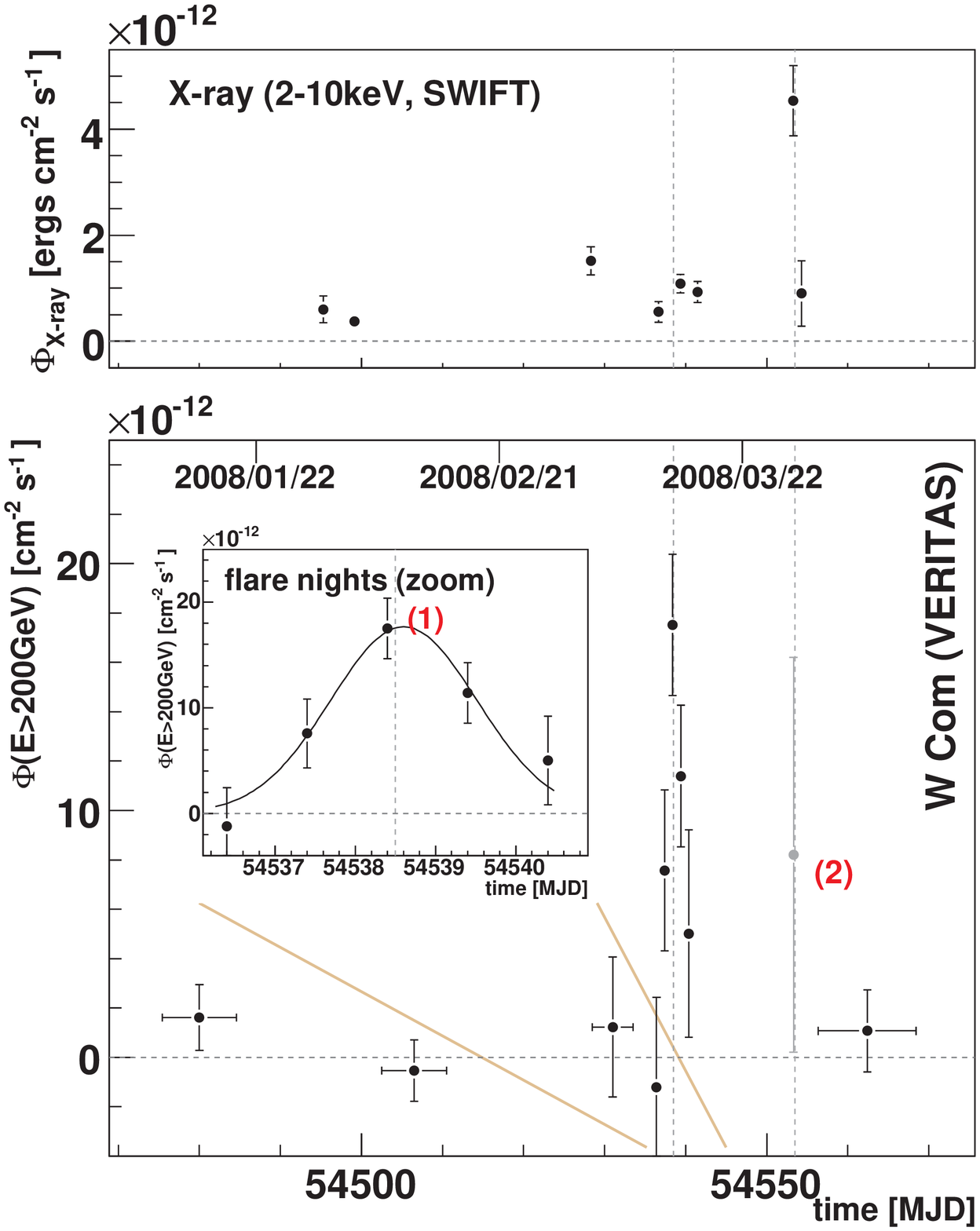} \label{fig:wcomlightcurve}}
  }
  
  \caption{\emph{Left Panel (\ref{fig:wcomskymap}):} Sky
    map of significances from W Com.  The excess
    $\sim2^\circ$ North of W Com corresponds to 1ES
    1218+304 and demonstrates the capability of VERITAS to
    detect sources at the edge of the FoV of the
    camera. \emph{Right Panel (\ref{fig:wcomlightcurve}):}
    The light curve of W Com is shown in the bottom
    section binned by observation period.  The inset is a
    night-by-night light curve of the four day flaring
    period.  The upper panel is the X-ray flux as measured
    by the \emph{Swift} telescope.}
  \label{fig:wcom}
\end{figure*}

Blazars are characterized by their double-humped spectral energy
distribution (SED) and are further classified according to the
location of the lower energy hump, usually interpreted as synchrotron
emission from relativistic electrons. Most of the $\sim$25 VHE blazars
detected by ground-based IACT are high-frequency peaked BL Lac (HBL)
objects.  VERITAS has discovered two HBL objects, 1ES 0806+524 and,
more recently RGB J0710+591.  In addition to these two HBL, VERITAS
has discovered two intermediate-frequency peaked BL Lac (IBL) objects,
3C 66A and W Com.  This contribution describes the discovery of VHE
$\gamma$-rays from these four blazars.  There are several other
contributions in this conference which will detail the VERITAS
observations of known blazars, including multiwavelength (MW)
observations and EBL studies (\cite{reyes.2009, imran.2009,
  grube.2009a, grube.2009b, hui.2009, maier.2009}).

\section{1ES 0806+524}

The discovery \cite{acciari.2009a} of the HBL 1ES 0806+524 was made
with a combination of observations from the commissioning phase of
VERITAS (30 hours) and full array time (35 hours). These observations
resulted in 245 excess events corresponding to a detection at the
6.3$\sigma$ level.  Figure \ref{fig:0806theta2} shows the distribution
of the squared angular distance between the shower direction and the
location of 1ES 0806+524 indicating that this source is point-like.
The photon spectrum shown in Figure \ref{fig:0806spectrum} is
characterized by a power law with photon index $3.6 \pm 1.0_{stat} \pm
0.3_{sys}$ between 300 and 700 GeV.  The integral flux above 300 GeV
is $2.2 \pm 0.5 \pm 0.4 \times 10^{-12} cm^{-2} s^{-1}$ corresponding
to 1.8\% of the Crab Nebula's flux.  Assuming absorption on the
infrared component of the EBL according to Franceschini \etal, 2008
\cite{franceschini.2008}, the de-absorbed spectrum is calculated to be
$2.8 \pm 0.5$.  The VERITAS data together with simultaneous
\emph{Swift} observations in the UV to optical to X-ray energies was
used to construct a spectral energy distribution (SED) of the
object. The SED is well fit by a pure synchrotron-self-Compton (SSC)
model \cite{bottcher.2002}. For more details on the multiwavelength
observations of this object see J.~Grube's talk in this conference
\cite{grube.2009a}.

\section{W Com}

\begin{figure}[!t]
  \centering
  \includegraphics[width=2.5in]{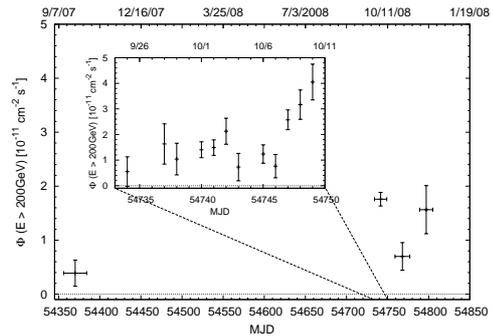}
  \caption{Light curve binned by observation period for 3C\,66A.
    These data indicate night-by-night variability for the second dark
    period but not within any of the other dark periods.  The inset
    shows the night-by-night lightcurve for the October observation
    period.}
  \label{fig:3C66Alightcurve}
\end{figure}

\begin{figure*}[!t]
  \centerline{\subfloat[Case I]{\includegraphics[width=3.2in]{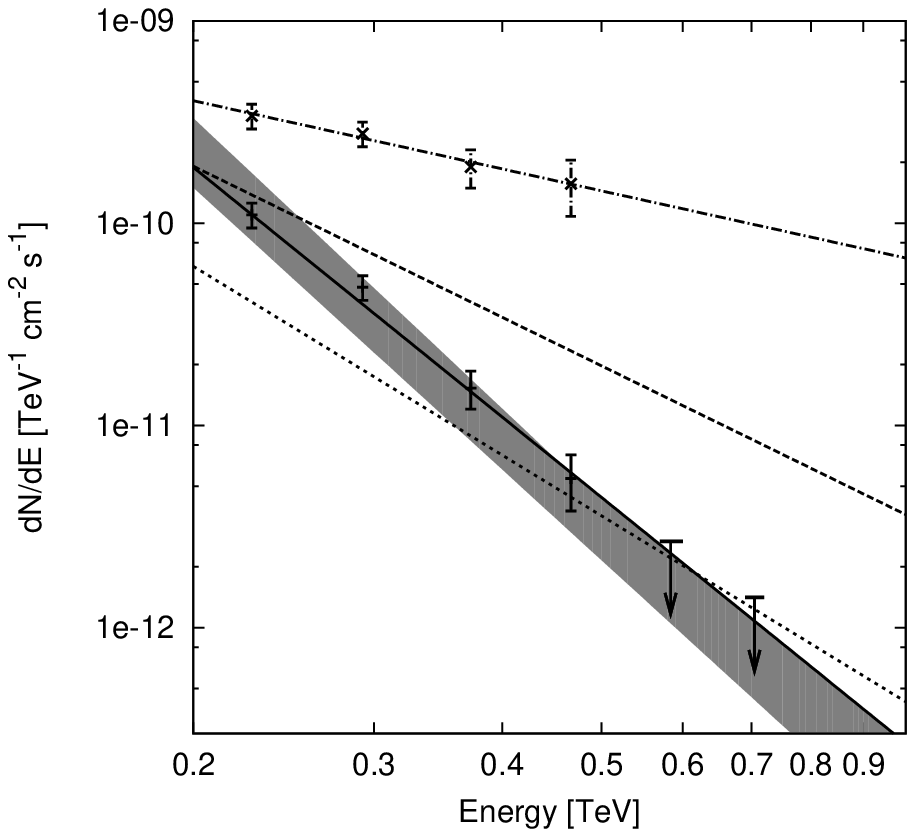} \label{fig:3C66Aspectrum}}
    \hfil
    \subfloat[Case II]{\includegraphics[width=2.5in]{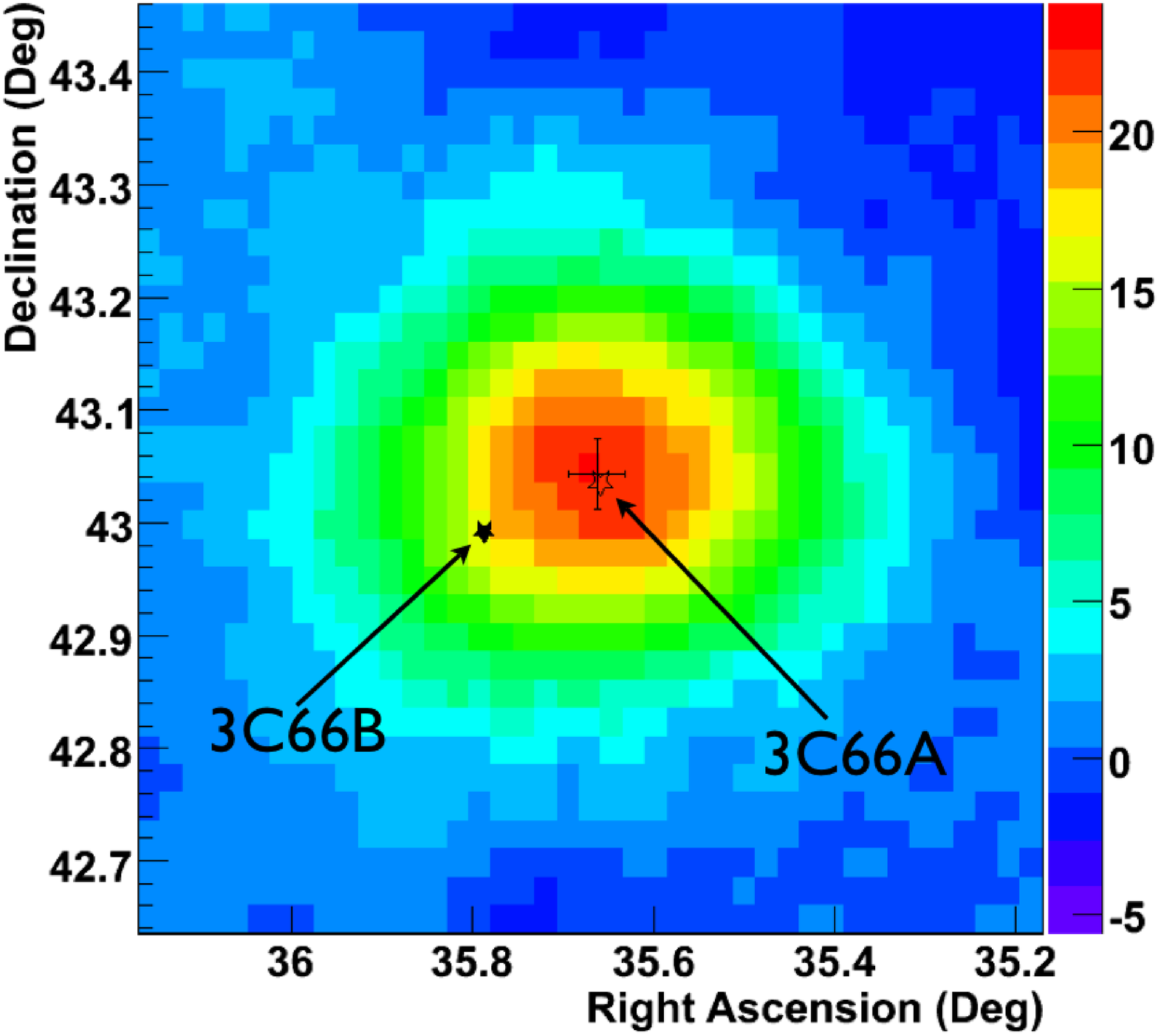} \label{fig:3C66Askymap}}
  }
  \caption{\emph{Left Panel (\ref{fig:3C66Aspectrum}):} The photon
    spectrum of 3C\,66A shown as solid points.  The spectrum is well
    fitted by a power law with index $\Gamma = 4.1 \pm 0.4_{stat} \pm
    0.6_{sys}$ (solid line). The shaded area outlines the systematic
    error in the photon spectral index. Using the model of
    Franceschini \etal\cite{franceschini.2008} and assuming a redshift
    of $z=0.444$, the de-absorbed spectral index is calculated to be
    $1.1 \pm 0.4$ showing that the very steep measured spectrum could
    be due to the distance of 3C\,66A.  The de-absorbed spectrum is
    shown as a dashed-dotted line and points.  The MAGIC
    spectrum\cite{aliu.2009} with index $\Gamma=3.1$ is shown as a
    dotted line.  The Crab Nebula's spectrum divided by 10 is also
    shown for comparison (dashed line). \emph{Right Panel
      (\ref{fig:3C66Askymap}):}Smoothed significance map of 3C\,66A.
    The location of 3C\,66A is shown as an open star and 3C\,66B as
    the closed star.  The cross is the fit to the excess VHE emission
    resulting in a localization of 2h 22m 41.6s $\pm$ 1.7s $\pm$ 6.0s,
    $43^o$ 02' 35.5'' $\pm$ 21'' $\pm$ 1'30''. Note that the bins are
    highly correlated due to an integration over angular space.}
  \label{fig:3C66A}
\end{figure*}

VERITAS discovered the first IBL in VHE $\gamma$-rays, W Com, in a
flaring state in March 2008 \cite{acciari.2008} (see Figure
\ref{fig:wcomskymap}).  W Com is a known $\gamma$-ray source detected
previously by EGRET in the 100 MeV - 10 GeV band
\cite{hartman.1999}. The flare, which lasted only four days (see
Figure \ref{fig:wcomlightcurve}) resulted in a detection at the
8$\sigma$ level.  The significance of the full data set is
4.9$\sigma$.  70\% of the total excess was from the four-day flaring
period and during the two brightest nights, W Comae was observed at
the 9\% Crab flux level. Using quasi-simultaneous multiwavelength
data, the SED can be reasonably modeled with a simple one-zone SSC but
this yields an extraordinarily low magnetic field value (B = 0.007 G).
A more natural set of fit parameters is obtained with an
external-Compton model indicating that the higher optical luminosity
is likely providing a seed population of photons for inverse-Compton
scattering. A second VHE flare was observed from W Comae in June 2008
wherein the flux was $\sim$3 times brighter than the previous flare
\cite{swordy.2008}.  For more details on this second flare, see
G.~Maier's contribution \cite{maier.2009} in this conference.  A
multiwavelength campaign was triggered during this time and will be
the subject of a future paper.  Interestingly, the VHE HBL 1ES
1218+304 can also be seen in the field-of-view (FoV) of the W Comae
observations which highlights the capabilities of off-axis
observations with VERITAS (see Figure \ref{fig:wcomskymap}).

\section{3C 66A}

The IBL 3C 66A was discovered by VERITAS in 33 hours of observations
resulting in a 21.2$\sigma$ detection (1791 excess events)
\cite{acciari.2009b}. Variability is seen on day time scales (see
Figure \ref{fig:3C66Alightcurve}) during a strong flare observed in
October, 2008. The observed spectrum is well-fit by a soft power law
with an index of $4.1 \pm 0.4_{stat} \pm 0.6_{sys}$.  The redshift of
3C 66A is uncertain.  It was measured to be 0.444 based on a single
line \cite{miller.1978} and there is a lower limit of 0.096
\cite{finke.2008}. If the measured value of 0.444 is correct, the
softness of the measured spectrum could be completely due to EBL
attenuation of VHE photons (see Figure \ref{fig:3C66Aspectrum}).
Using the model of Franceschini \etal\cite{franceschini.2008} and
assuming a redshift of $z=0.444$, the de-absorbed spectral index is
calculated to be a very hard $1.1 \pm 0.4$.

3C 66A lies only $0.12^\circ$ from the radio galaxy 3C 66B and care
must be taken in identifying the source of $\gamma$-rays from this
region.  Recently, the MAGIC collaboration reported the detection of
VHE gamma-rays from the region in observations carried out in 2007 and
claimed marginal evidence (at the 85\% confidence level) for
association with 3C 66B \cite{aliu.2009}.  However, the VERITAS data
exclude the position of 3C 66B at the 4.3$\sigma$ level (see Figure
\ref{fig:3C66Askymap}).  In addition to this, the measured VERITAS
spectrum does not agree with that measured by MAGIC.  One explanation
of the discrepancy in source localization is that 3C\,66B must have
been considerably brighter in 2007 than 2008, and similarly 3C 66A
must have been considerably brighter in 2008 than 2007.

Recently the LAT instrument on the Fermi satellite also observed
bright emission in the MeV-GeV range at a level higher than that
reported by EGRET.  In addition, observations of the blazar were made
by the \emph{Swift} and \emph{Chandra} satellites.  Details of these
multiwavelength observations will be the subject of a future paper and
are highlighted in L.~Reyes' contribution \cite{reyes.2009}.

\section{RGB J0710+591}

\begin{figure}[!t]
  \centering
  \includegraphics[width=2.5in]{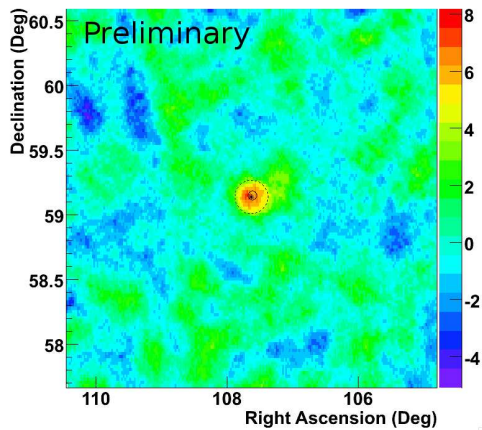}
  \caption{Preliminary significance skymap of the region centered on
    RGB J0710+591.  The star is the location of the HBL RGB J0710+591.
    The solid circle is the fit to the position of the $\gamma$-ray
    excess including the statistical and systematic error.  The dashed
    circle is the integration region.}
  \label{fig:RGBJ0710skymap}
\end{figure}

RGB J0710+591 was only recently discovered to emit VHE emission and is
the fourth VERITAS blazar discovery \cite{ong.2009}.  Located at a
redshift of 0.125, VERITAS observed RGB J0710 for $\sim20$ hours
resulting in a $>6\sigma$ detection from 140 $\gamma$-rays
corresponding to $\sim2\%$ of the Crab Nebula's flux.  A future
publication will describe the VERITAS observations as well as
multiwavelength observations from the \emph{Swift}, \emph{Chandra} and Fermi
satellites.  A preliminary skymap is shown in Figure
\ref{fig:RGBJ0710skymap}.  The preliminary observed spectrum can be
described as a power law with photon index $\Gamma=2.8 \pm 0.3_{stat}
\pm 0.3_{sys}$.  While not as hard as 1ES 0229+200 which placed the
strongest constraints to date on the density of the EBL in the
mind-infrared band, the moderately high distance and relatively hard
spectrum of RGB J0710+591 confirms the constraints presented in
\cite{aharonian.2007}.

\begin{table*}[th]
  \caption{VERITAS Blazar Discoveries.}
  \label{tab:blazars}
  \centering
  \begin{tabular}{|c|c|c|c|c|c|ccc|c|}
    \hline
    Source & Type & Redshift & Exposure & Excess & Significance &\multicolumn{3}{|c|}{Integral
      Flux} & Spectral Index\\
    & & & [hours] & & [$\sigma$] & [$10^{-11} cm^{-2} s^{-1}$] & [\% crab] & [GeV] & [$\Gamma$]\\
    \hline 
    W Com       & IBL & 0.102 & 39.5 & 111 & 4.9\footnotemark & 1.99 & 9.0 & $>200$ & $3.8\pm0.4_{stat}\pm0.3_{sys}$ \\
    RGB J0710+591 & HBL & 0.125 & 22.4 & 141 & 6.3 & & & & $2.7\pm0.3_{stat}\pm0.3_{sys}$ \\
    1ES 0806+524  & HBL & 0.138 & 65.0 & 245 & 6.3 & 0.22 & 1.8 & $>300$ & $3.6\pm1.0_{stat}\pm0.3_{sys}$ \\
    3C 66A        & IBL & 0.444 & 32.8 & 1791 & 21.2 & 1.30 & 6.0 & $>200$ & $4.1\pm0.4_{stat}\pm0.6{sys}$ \\
    \hline
  \end{tabular}
\end{table*}

\section{Conclusions}  

To date, VERITAS has detected more than a dozen blazars and has
discovered VHE emission from four blazars: 1ES 0806+524, RGB J0710+591
W Com and 3C 66A (see Table \ref{tab:blazars}).  The first two of
these are HBL objects while the latter two are IBL objects.  The
discovery of 1ES 0806+524 highlights the capabilities of VERITAS to
detect low-flux objects while the detection of the two IBL objects
opens a new window into the study of blazar populations in the VHE
regime.  Additionally, the detection of 1ES 1218+304 in the FoV of W
Com provides an example of the ability to detect objects at the edge
of VERITAS' FoV.  Future studies of the moderately distant RGB
J0710+591 might provide interesting constrains on the EBL.

\newpage

\section{Acknowledgments}  
This research was supported by grants from the U.S. Department of
Energy, the U.S. National Science Foundation and the Smithsonian
Institution, by NSERC in Canada, by Science Foundation Ireland and by
STFC in the UK.

\footnotetext{Significance for the full data set.  The significance
  of the flaring event was $8~\sigma$.}


\begin{thebibliography}{99}

\bibitem{ong.2009} R.~Ong for the VERITAS Collaboration. \emph{VERITAS
    Discovery of VHE Gamma-Ray Emission from BL Lac object RGB
    J0710+591}, The Astronomer's Telegram, \#1941.

\bibitem{weekes.2002} T.~C.~Weekes \etal, APh, 17 (2002), 221

\bibitem{benbow.2009} W.~Benbow for the VERITAS Collaboration,
  \emph{The VERITAS Blazar Key Science Project}, Proceedings of the
  31\MakeLowercase{$^{st}$} ICRC, {\L}\'{o}d\'{z} 2009

\bibitem{reyes.2009} L.~Reyes for the VERITAS Collaboration,
  \emph{Simultaneous Observations of flaring gamma-ray blazar 3C 66A
    with Fermi and VERITAS}, Proceedings of the
  31\MakeLowercase{$^{st}$} ICRC, {\L}\'{o}d\'{z} 2009

\bibitem{imran.2009} A.~Imran for the VERITAS Collaboration,
  \emph{Connecting the EBL with the Hard Spectra of VHE Blazars: New
    Results from VERITAS}, Proceedings of the
  31\MakeLowercase{$^{st}$} ICRC, {\L}\'{o}d\'{z} 2009

\bibitem{grube.2009a} J.~Grube for the VERITAS Collaboration,
  \emph{Highlights of recent multiwavelength observations of VHE
    blazars with VERITAS}, Proceedings of the
  31\MakeLowercase{$^{st}$} ICRC, {\L}\'{o}d\'{z} 2009

\bibitem{grube.2009b} J.~Grube for the VERITAS Collaboration,
  \emph{Detailed five day flaring observations of Mrk 421 with Suzaku
    and VERITAS in May 2008}, Proceedings of the
  31\MakeLowercase{$^{st}$} ICRC, {\L}\'{o}d\'{z} 2009

\bibitem{hui.2009} M.~Hui for the VERITAS Collaboration, \emph{VERITAS
    observations of M87 from 2007 to present}, Proceedings of the
  31\MakeLowercase{$^{st}$} ICRC, {\L}\'{o}d\'{z} 2009

\bibitem{maier.2009} G.~Maier for the VERITAS Collaboration,
  \emph{Multiwavelength observations of a TeV-Flare from W Comae},
  Proceedings of the 31\MakeLowercase{$^{st}$} ICRC, {\L}\'{o}d\'{z}
  2009

\bibitem{acciari.2009a} V.~Acciari \etal, ApJL, 690 (2009) L126-L129

\bibitem{franceschini.2008} A.~Franceschini, G.~Rodighiero, \&
  M.~Vaccari, A\&A, 487 (2008) 837

\bibitem{bottcher.2002} M.~B\"{o}ttcher \& J.~Chiang, ApJ, 581 (2002)
  127

\bibitem{acciari.2008} V.~Acciari \etal, ApJL, 684 (2008) L73-L77

\bibitem{hartman.1999} R.~C.~Hartman \etal, ApJS, 123 (1999) 79

\bibitem{swordy.2008} S.~Swordy for the VERITAS
  Collaboration. \emph{TeV Outburst from W Comae}, The Astronomer's
  Telegram, \#1565.

\bibitem{acciari.2009b} V.~Acciari \etal, ApJL, 693 (2009) L104-L108

\bibitem{miller.1978} J.~S.~{Miller}, H.~B.~{French} \&
  S.~A.~{Hawley}, BL Lac Objects (1978) 176

\bibitem{finke.2008} J.~D.~{Finke}, J.~C.~{Shields}, M.~{B{\"o}ttcher}
  \& S.~{Basu}, AAP, 477 (2008), 513

\bibitem{aliu.2009} E.~Aliu \etal, ApJL, 692 (2009), L29

\bibitem{aharonian.2007} F.~Aharonian \etal, A\&A, 475 (2007), L90L13

\end{thebibliography}
\end{document}